\documentclass[prb,twocolumn,showpacs]{revtex4}% Physical Review B

\usepackage{graphicx,epsf}% Include figure files
\usepackage{bm}% bold math

\begin{document}

\title{The Coulomb Blockade in Quantum Boxes}
\smallskip

\author{Eran Lebanon,$^1$ Avraham Schiller,$^1$ and Frithjof B.~Anders$^2$}
\affiliation{$^1$Racah Institute of Physics, The Hebrew University,
                 Jerusalem 91904, Israel \\
             $^2$Institut f\"ur Festk\"orperphysik, TU Darmstadt,
	         64289 Darmstadt, Germany}

\begin{abstract}
The charging of a quantum box connected to a lead by a single-mode
point contact is solved for arbitrary temperatures, tunneling amplitudes,
and gate voltages, using a variant of Wilson's numerical renormalization
group. The charge inside the box and the capacitance of the junction
are calculated on equal footing for all physical regimes, including
weak tunneling, near perfect transmission, and the crossover regime
in between. At the charge plateaus, perturbation theory is found to
break down at fairly small tunneling amplitudes. Near perfect transmission,
we confirm Matveev's scenario for the smearing of the Coulomb-blockade
staircase. A surprising reentrance of the Coulomb-blockade staircase
is found for large tunneling amplitudes. At the degeneracy points,
we obtain two-channel Kondo behavior directly from the Coulomb-blockade 
Hamiltonian, without the restriction to two charge configurations
or the introduction of an effective cutoff.
\end{abstract}

\maketitle
% General introduction
The Coulomb blockade~\cite{CB1,CB2} is one of the fundamental
phenomena in mesoscopic physics. When a quantum box, either a
small metallic grain or a large semiconducting quantum dot, is
coupled by weak tunneling to a lead, its charging is governed
by the finite energy barrier $E_C$ for adding a single electron
to the box. This gives rise to the well-known Coulomb-blockade
staircase for the charge of the box as a function of gate
voltage.~\cite{CB1,CB2} Increasing the coupling to the lead smears
the Coulomb staircase, as thermal fluctuations do at $k_B T>E_C$.

% more specific introduction 
A large diversity of theoretical approaches have been applied
to the Coulomb blockade, ranging from perturbation 
theory~\cite{SZ90,GM90,Matveev91,Grabert_et_al} and diagrammatic
techniques,~\cite{SS94,GZ94,LSZ01} to renormalization-group
treatments~\cite{GM90,Matveev91,KS98,FSZ94_95,HZ97}
and Monte Carlo simulations.~\cite{FSZ94_95,HZ97,WEG97,HSZ99}
However, to date there is no single unified approach encompassing 
all regimes of the Coulomb blockade. The reason being that 
different starting points are required for describing the
different physical regimes of the Coulomb blockade. For weak
tunneling and $k_B T \ll E_C$, the charge inside the box is
essentially quantized
at the charge plateaus. Hence, one can start from a well-defined
charge configuration and apply the perturbation theory in the
tunneling amplitude. This approach collapses at the degeneracy
points between adjacent charge plateaus, where strong charge
fluctuations give rise to exotic many-body physics in the form
of the two-channel Kondo effect.~\cite{Matveev91,CZ98} For
large transmission, the Coulomb-blockade staircase is smeared 
out and the notion of well-defined charge configurations
breaks down.

% our achivements.
In this paper, we devise such a unified approach for all temperatures
and parameter regimes of the Coulomb blockade using Wilson's numerical 
renormalization-group method.~\cite{Wilson75} Focusing on single-mode 
point contacts, we accurately calculate the charge and the capacitance
of the quantum box, for arbitrary gate voltages, temperatures, and
tunneling amplitudes. These quantities were recently measured by
Berman {\em et al}. for single-mode point contacts,~\cite{BZAS99}
revealing signatures of the two-channel Kondo effect. For weak
tunneling, we recover the results of the perturbation theory at the
charge plateaus. However, we find that the perturbation theory breaks 
down at fairly small tunneling amplitudes. Close to perfect transmission,
we find good agreement with Matveev's analysis of a related
model,~\cite{Matveev95} which in turn breaks down as one departs
from perfect transmission. Upon further increasing the tunneling
matrix element, there is a surprising reentrance of the 
Coulomb-blockade staircase. At the degeneracy points, we obtain
two-channel Kondo behavior directly from the Coulomb-blockade 
Hamiltonian, without the restriction to two charge configurations
or the introduction of an effective cutoff. Indeed, the effective
cutoff entering Matveev's two-charge-configuration
model~\cite{Matveev91} is found to depend not only on the
charging energy $E_C$, but also on the tunneling strength.

% model
The coupled quantum-box--lead system is conventionally modeled 
by the Hamiltonian 
\begin{eqnarray}
{\cal H} & = & \sum_{\alpha=L,B} \sum_{k\sigma}
                     \epsilon_{\alpha k} 
                     c^{\dagger}_{\alpha k\sigma}c_{\alpha k\sigma}
           + \; \frac{\hat Q_B^2}{2C_0}+V_B \hat Q_B
\label{initial_Hamiltonian}
\\
&& + \;
      t \sum_{kk' \sigma}
      \left( 
            c^{\dagger}_{Lk\sigma}c_{B k'\sigma} + {\rm H.c.}
      \right) ,
\nonumber 
\end{eqnarray}
where $c^{\dagger}_{L k \sigma}$ ($c^{\dagger}_{B k \sigma}$) creates 
a lead (box) electron with momentum $k$ and spin $\sigma$; $t$ is the 
tunneling matrix element, taken for simplicity to be momentum 
independent; and $\epsilon_{L k}$ ($\epsilon_{B k}$) are the single-particle 
levels in the lead (box), which are assumed to be continuous (dense 
energy levels). The excess electrical charge inside the box, 
\begin{equation}
\hat Q_B =
-e\sum_{k\sigma} [c^{\dagger}_{Bk\sigma}c_{Bk\sigma}-\theta(-\epsilon_{Bk})]
\end{equation}
($-e$ being the electron charge), is controlled by the capacitance of
the box, $C_0$, and by the external component of the electrostatic
potential in the box, $V_B$.

%one large pragraph: mapping
The numerical renormalization group~\cite{Wilson75} (NRG) is a
nonperturbative approach for treating quantum impurity problems.
At the center of this approach is a logarithmic energy discretization 
of the conduction band around the Fermi level, designed to capture
all logarithmic divergences of the problem. Using a unitary
transformation, the conduction band is mapped onto
a semi-infinite chain with the impurity coupled to the open end,
and a hopping matrix element that decreases exponentially along the
chain. A direct application of this approach to the Coulomb-blockade
Hamiltonian is doomed to fail, since the Coulomb interaction equally
couples all sites along the semi-infinite chain. It does not decay
as the hopping matrix elements. To circumvent this problem, we resort
to a mapping used by Schoeller and co-workers.~\cite{SS94,KS98}
Explicitly, we map the Hamiltonian of Eq.~(\ref{initial_Hamiltonian})
onto
\begin{eqnarray}
{\cal H} & = &\sum_{\alpha=L,B} \sum_{k\sigma}
              \epsilon_{\alpha k}
              c^{\dagger}_{\alpha k\sigma}c_{\alpha k\sigma} 
           + \frac{{\hat Q}^2}{2C_0} + V_B {\hat Q}
\label{mapped_Hamiltonian}
\\
&& + \; t\sum_{kk'\sigma}
             \left\{
	             {\hat Q}^{-} c^{\dagger}_{L k\sigma} c_{B k'\sigma}
		     + {\rm H.c.}
             \right\} ,
\nonumber 
\end{eqnarray}
where
\begin{eqnarray}
{\hat Q}^{\pm} = \sum_n |n \pm 1\rangle\langle n | \ \ \ {\rm and} \ \ \
{\hat Q}       = -e\sum_n n |n \rangle\langle n |
\end{eqnarray}
are new collective charge operators. Strictly speaking, the above
mapping requires the constraint ${\hat Q}= \hat Q_B$. However, this
constraint can be relaxed in the continuum limit, when the
dynamics of $\hat Q$ in Eq.~(\ref{mapped_Hamiltonian}) is
insensitive to the precise number of conduction electrons in
the bands. Hereafter, we regard ${\hat Q}$ as an independent
degree of freedom.

\begin{figure}[tb]
\centerline{
\includegraphics[width=75mm]{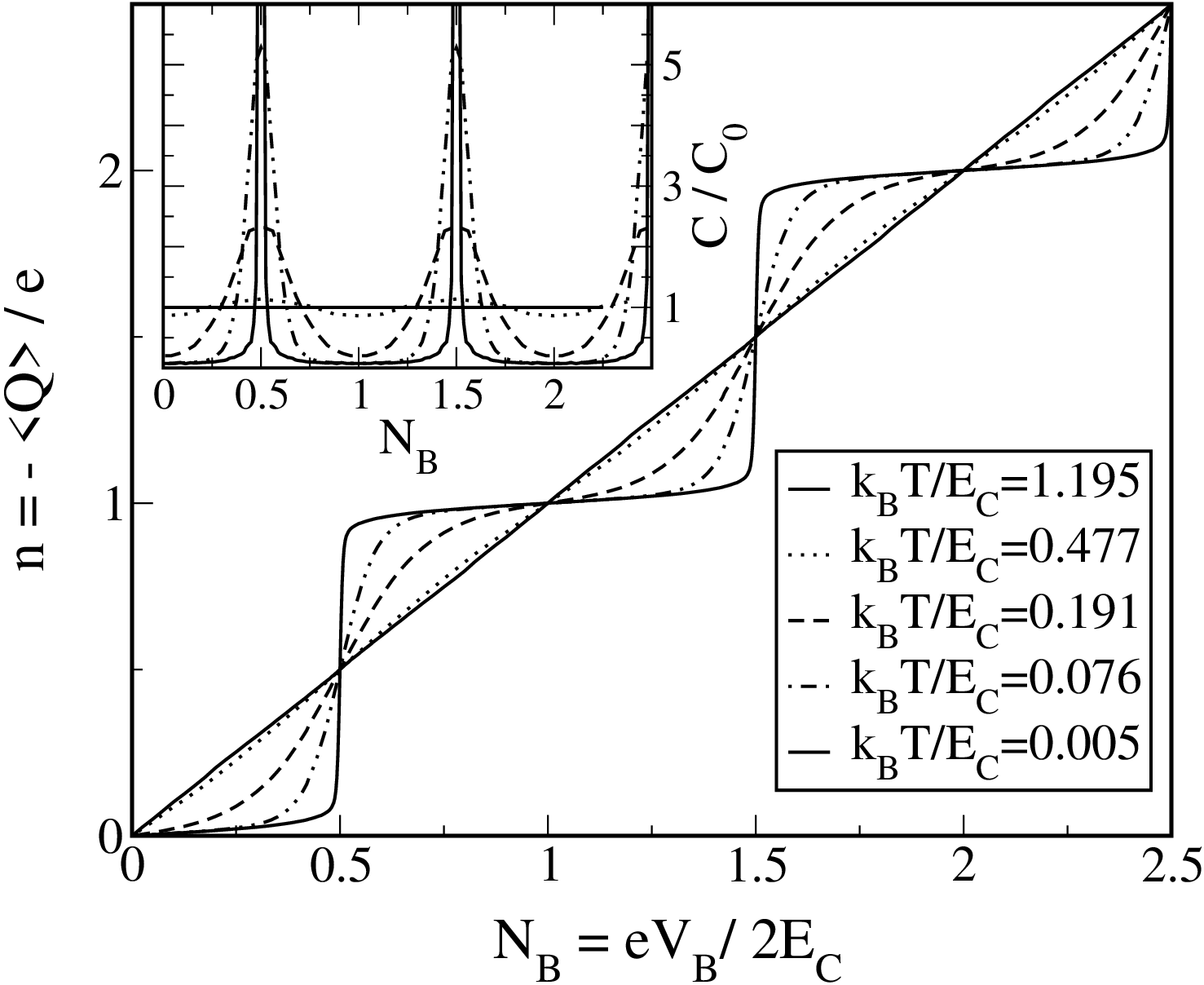}
}\vspace{0pt}
\caption{The excess charge inside the box versus $V_B$, for
         $t\rho=0.1$ and different temperatures. Here,
	 $E_C/D = 0.01$, $\Lambda = 2.5$, and
	 $N_{\rm s} = 2000$. At temperatures comparable to
	 the charging energy, the charge curve is essentially
	 linear. A sharp Coulomb-blockade staircase emerges for
         $k_B T \ll E_C$. Inset: The capacitance versus $V_B$,
         for the same temperatures as plotted in the main
         panel. For $k_B T > E_C$, the capacitance is constant
         and equal to $C_0 = e^2/2E_C$. For $k_B T \ll E_C$,
	 sharp capacitance peaks develop at the degeneracy points.}
\label{fig:fig1}
\end{figure}

Equation~(\ref{mapped_Hamiltonian}) describes two noninteracting
conduction bands, coupled to an impurity with an infinite number
of parabolically dispersed energy levels. Hence, it can be treated
using the NRG. We solved this model using a common constant density
of states $\rho$ for the lead and the box,~\cite{DOS} and a common
bandwidth $D$. The NRG method is controlled by the logarithmic
discretization parameter $\Lambda$ and by the number of states
retained $N_{\rm s}$.~\cite{Wilson75} 

% General results
Figure~\ref{fig:fig1} shows the expected Coulomb-blockade staircase
at different temperatures, for the weak tunneling $t\rho=0.1$. At
temperatures above the charging energy $E_C=e^2/(2C_0)$, one recovers
a linear (classical) charge curve. As the temperature is reduced below
the charging energy, a staircase structure emerges, with a period equal
to twice the charging energy. For $k_B T \ll E_C$, one is left with
sharp charge steps, with abrupt transitions between the charge
plateaus at half-integer values of $N_B=eV_B/2E_C$.

A complimentary picture is provided by the capacitance,
$C = -d\langle \hat Q \rangle/dV_B$,
recently measured by Berman {\em et al}.~\cite{BZAS99} As seen
in the inset of Fig.~\ref{fig:fig1}, the capacitance $C$ is equal to
the classical capacitance $C_0$, for $k_B T \gtrsim E_C$. However,
$C$ acquires a strong dependence on $N_B$, for $k_B T \ll E_C$.
While $C$ is significantly reduced at the charge plateaus, near
the degeneracy points (i.e., near half-integer values of $N_B$)
it is dramatically enhanced. As will be discussed below, $C$
diverges logarithmically at the degeneracy points as $T \to 0$,
in accordance with the two-channel Kondo effect predicted by
Matveev.~\cite{Matveev91}

% Plateau
At the charge plateaus, perturbation theory in $t\rho$ is commonly used
for weak tunneling. To second order in $t\rho$, the zero-temperature
excess number of electrons inside the box, $n=-\langle \hat Q \rangle/e$,
is given by
\begin{eqnarray}
n &=& 2(t\rho)^2
       \left\{
               \ln \left( \frac{1+2N_B}{1-2N_B} \right)
       \right.
\label{perturbative_expression}
\\
&& +   \left.
               2\ln \left( \frac{1-2N_B+d}{1+2N_B+d} \right) +
                \ln \left(\frac{1+2N_B+2d}{1-2N_B+2d}\right)
       \right\}.
\nonumber
\end{eqnarray}
Here, we have considered particle-hole symmetric bands, and retained a 
finite ratio $d=D/E_C$. In the wide-band limit, $d\rightarrow \infty$, the 
last two terms drop in Eq.~(\ref{perturbative_expression}), and one 
recovers the conventional expression (see, e.g.,
Ref.~\onlinecite{Grabert_et_al}).

\begin{figure}[tb]
\centerline{
\includegraphics[width=75mm]{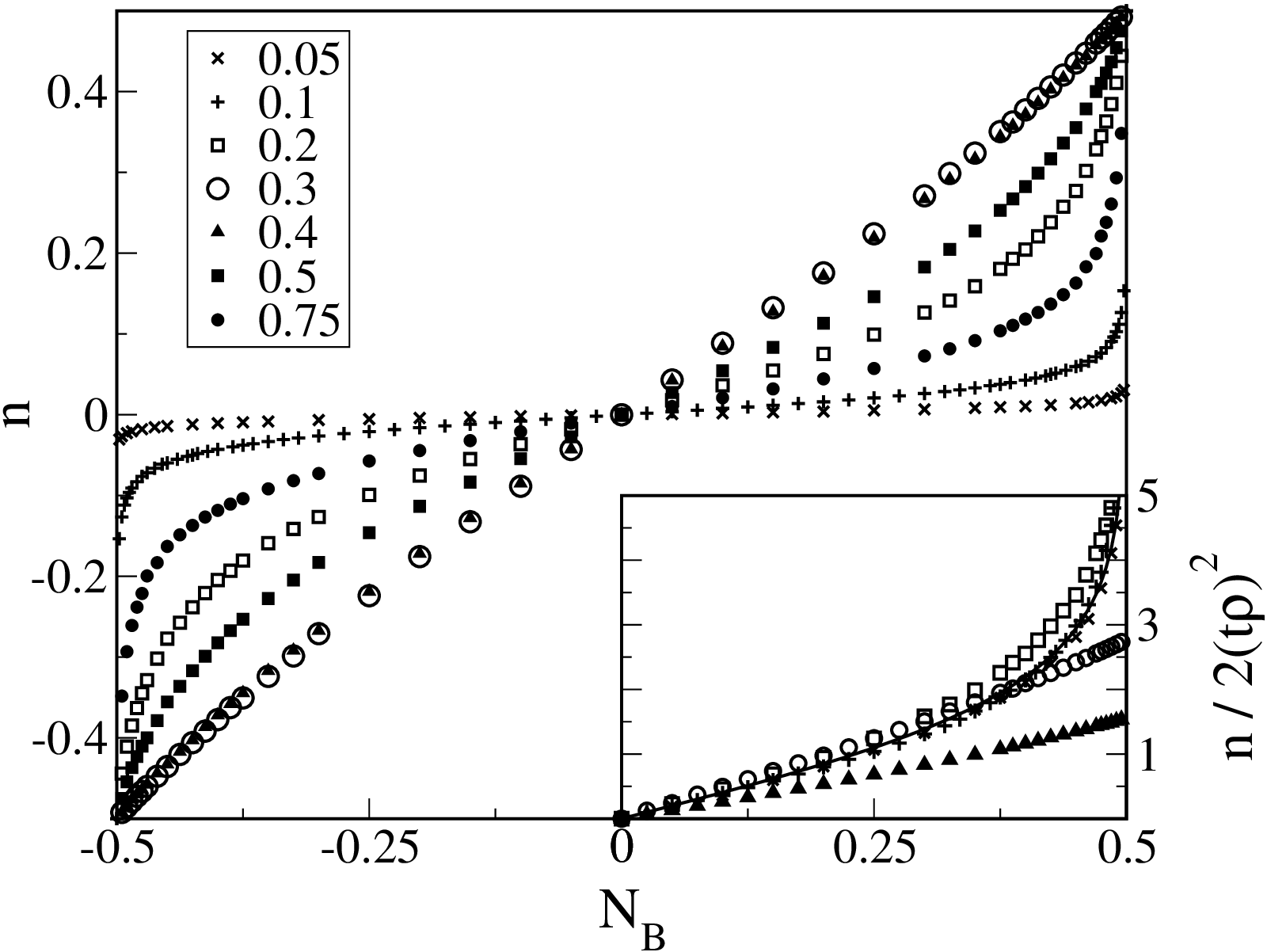}
}\vspace{0pt}
\caption{Evolution of the zero-temperature charging curve with
         increasing $t\rho$. Here, $E_C/D = 0.01$, $\Lambda = 2.5$,
	 and $N_{\rm s} = 2000$. The values of $t\rho$ are
	 specified in the legend. At very weak coupling, there
	 is a sharp Coulomb-blockade staircase. As demonstrated
	 in the inset, good agreement is obtained with the
	 perturbative expression for $n/2(t\rho)^2$ [see
	 Eq.~(\ref{perturbative_expression})], which is plotted
	 for comparison by the solid line. Significant deviations
	 from the perturbation theory develop for $t\rho \agt 0.2$.
	 Specifically, the Coulomb-blockade staircase is completely
	 washed out for $t\rho = 0.3$, marking the breakdown of
	 the perturbation theory. For $t\rho > 0.4$, there is a
	 reentrance of the Coulomb-blockade staircase.}
\label{fig:fig2}
\end{figure}

The inset of Fig.~\ref{fig:fig2} compares between the NRG and the
perturbative expression of Eq.~(\ref{perturbative_expression}). At
very weak tunneling, $t\rho=0.05$ and $0.1$, the NRG results coincide
with the perturbative expression. The good agreement extends to all
gate voltages except the vicinity of the degeneracy points. However,
significant deviations from Eq.~(\ref{perturbative_expression})
are seen at fairly weak coupling, $t\rho \approx 0.2-0.3$. Indeed,
the Coulomb-blockade staircase is completely washed out for
$t\rho = 0.3$ (main panel of Fig.~\ref{fig:fig2}), indicating the
breakdown of the perturbation theory. Moreover, there is a reentrance
of the Coulomb-blockade staircase for $t\rho > 0.4$, which obviously
is not captured by the perturbation theory in $t\rho$. 

The rapid breakdown of the perturbation theory is related to the
approach to perfect transmission. For $E_C=0$ and particle-hole
symmetric bands, the single-particle transmission coefficient at
the Fermi level is given by
\begin{equation}
{\cal T} = 4 \frac{ (\pi t \rho)^2 }
                  { \left[ 1+(\pi t \rho)^2\right]^2 } \; .
\label{transmission_coefficient}
\end{equation}
For $\pi t\rho=1$, ${\cal T}$ reaches perfect transmission and is
no longer perturbative in $t\rho$. That $\cal T$ rather than $t\rho$
is the relevant physical parameter for $E_C \ll D$ is evident from
Fig.~\ref{fig:fig2}. The Coulomb-blockade staircase is washed out
for $t\rho \sim 0.3-0.4$, when ${\cal T}$ of
Eq.~(\ref{transmission_coefficient}) approaches unity. The reentrance
of the Coulomb-blockade staircase for $t\rho > 0.4$ stems from the
reduction of the single-particle transmission coefficient with increasing
$t\rho>1/\pi$. It should be emphasized, however, that this picture is
only valid for $E_c\ll D$. Upon increasing $E_c$, the charge curves
cannot be exclusively parametrized by the transmission coefficient of
Eq.~(\ref{transmission_coefficient}). In particular, the
washing out of the Coulomb-blockade staircase and its subsequent
reentrance are pushed to higher values of $t\rho$, reflecting the
limitations of the single-particle picture. 

\begin{figure}[tb]
\centerline{
\includegraphics[width=75mm]{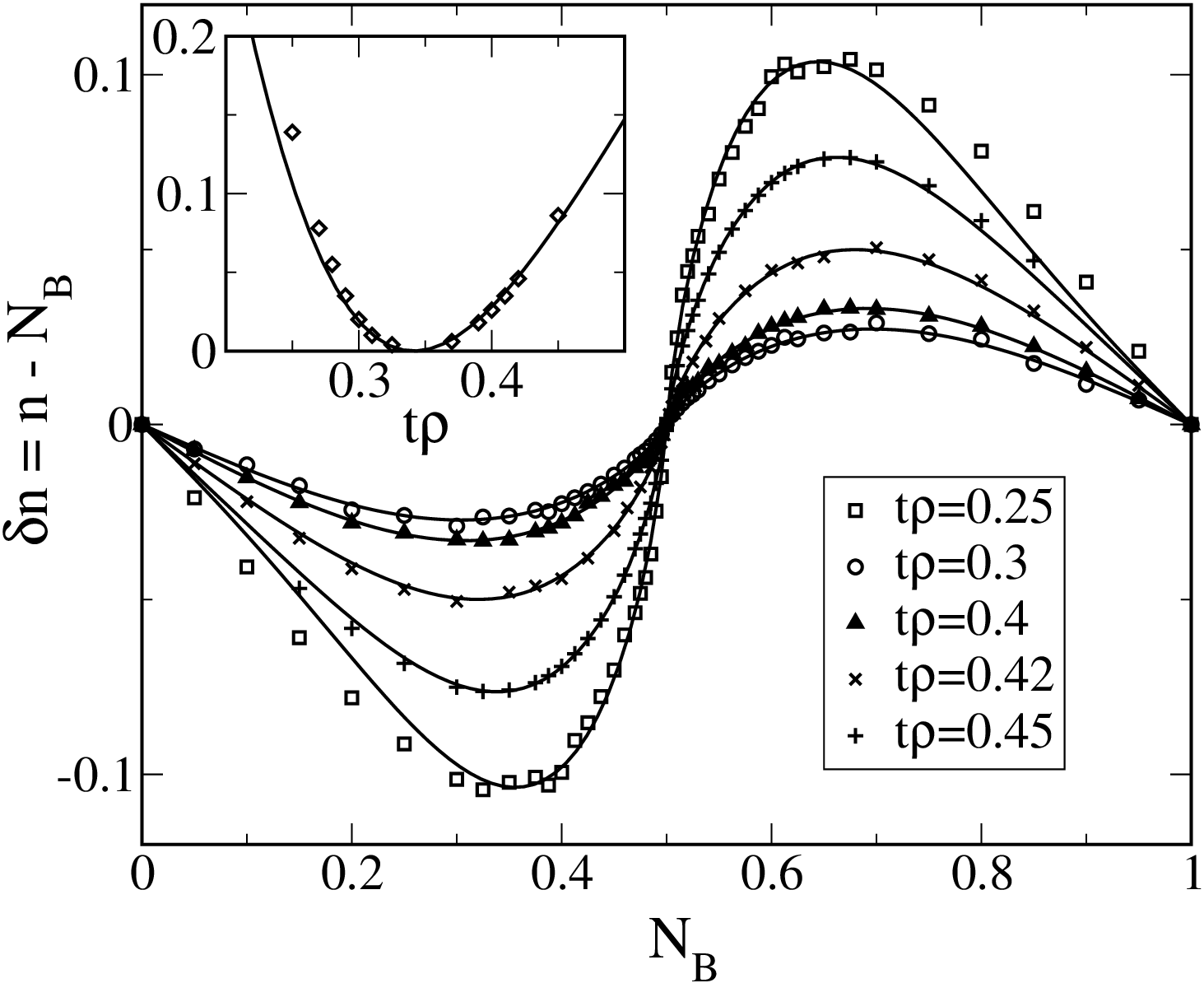}
}\vspace{0pt}
\caption{Comparison of the NRG (symbols) and Eq.~(\ref{bosonization}),
         for the zero-temperature charge curve near perfect transmission.
	 Here, $E_C/D = 0.01$, $\Lambda = 2.5$, and $N_{\rm s} = 2000$.
	 The horizontal axis, $\delta n = n - N_B$, measures the
	 deviation from linearity of the excess number of electrons
	 inside the box. The
	 solid lines are fits to Eq.~(\ref{bosonization}), using the
	 single fitting parameter $|r|^2$. Inset: A plot of the
	 extracted values of $|r|^2$ versus $t\rho$. The solid line
	 shows $1 - {\cal T}$ of Eq.~(\ref{transmission_coefficient})
	 with $t\rho \to t\rho - 0.023$.}
\label{fig:fig3}
\end{figure}

%strong coupling 
Near perfect transmission, it was predicted by Matveev that the
zero-temperature excess number of electrons inside the box is
given by~\cite{Matveev95}
\begin{eqnarray}
n = N_B + \frac{2\gamma|r|^2}{\pi^2}
          \ln\left[
	            e|r|^2\cos^2(\pi N_B)
	  \right] \sin(2\pi N_B) \, .
\label{bosonization}
\end{eqnarray}
Here, $|r|^2=1-{\cal T}$ is the reflectance, $\gamma$ is equal to
$e^C$, and $C \approx 0.5772$ is Euler's constant. This result
was obtained within a one-dimensional model, especially designed
for ${\cal T} \approx 1$. This regime of large transmission was
never studied using a tunneling Hamiltonian.

Figure~\ref{fig:fig3} shows a comparison of our NRG results for
$\delta n = n - N_B$ and Eq.~(\ref{bosonization}), using the
single fitting parameter $|r|^2$. There is a good agreement
between the two curves, particularly for $0.3 \leq t\rho \leq 0.4$.
Moreover, the extracted values of $|r|^2$ closely trace $1 - {\cal T}$
in this range of $t\rho$ (see inset of Fig.~\ref{fig:fig3}),
apart from a slight shift which is likely due to the finite value
of $E_C/D$ used. For $t\rho=0.25$ and $t\rho=0.45$, small systematic
deviations from Eq.~(\ref{bosonization}) develop in the middle
of the plateaus, near integer values of $N_B$. Further away from
perfect transmission, e.g., for $t\rho=0.2$ and $t\rho=0.5$ 
(not shown), there are large deviations from the analytic expression, 
signaling the breakdown of Matveev's expression. This clearly
shows that one can still use the tunneling Hamiltonian of
Eq.~(\ref{initial_Hamiltonian}) near perfect transmission, provided
that the tunneling processes are summed to all orders.

% 2CK
Of particular interest is the behavior near the degeneracy points,
where the perturbation theory diverges logarithmically even for small
values of $t\rho$. For weak tunneling and $k_B T\ll E_C$, the charge
fluctuations in this regime were mapped by Matveev~\cite{Matveev91}
onto the planar two-channel Kondo model, by retaining only
the two lowest-lying charge configurations inside the box.
The capacitance at the degeneracy points was predicted to diverge
logarithmically with $T\to 0$, in accordance with the logarithmic
divergence of the magnetic susceptibility in the two-channel Kondo
effect.~\cite{CZ98} This scenario was never confirmed directly
for the Hamiltonian of Eq.~(\ref{initial_Hamiltonian}).

\begin{figure}[tb]
\centerline{
\includegraphics[width=75mm]{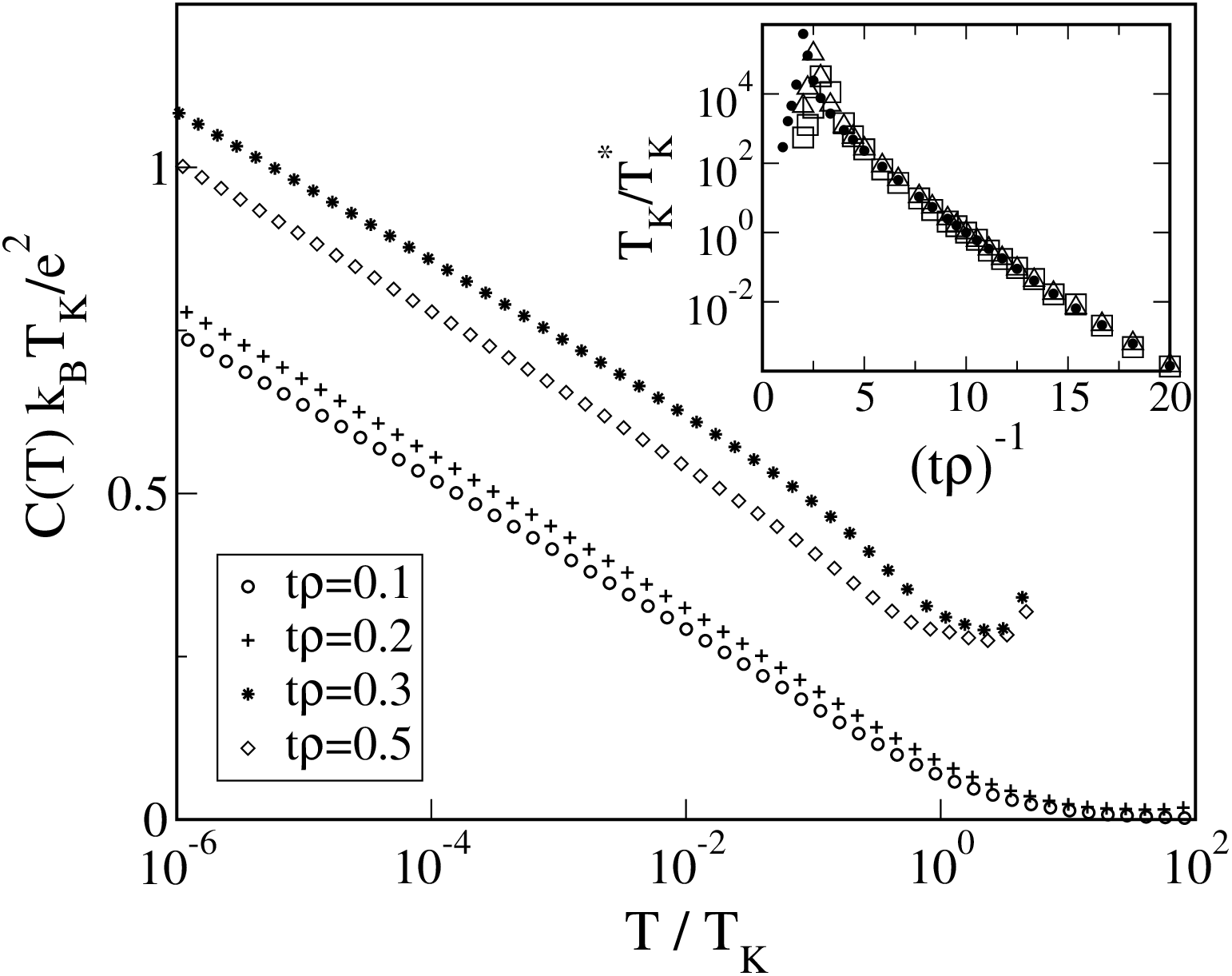}
}\vspace{0pt}
\caption{The capacitance at the degeneracy points as a function of
         temperature, for $E_C/D = 0.25$ and different values of $t\rho$.
	 Here, $\Lambda = 2$ and $N_s = 2000$. All curves show the
	 characteristic logarithmic temperature dependence of the
	 two-channel Kondo effect. Using Eq.~(\ref{Tk_def}), we
	 extract the Kondo temperature $T_K$ from the slope of the
	 logarithmically diverging component of $C(T)$. Inset:
	 $T_K/T_K^{\ast}$ versus $(t\rho)^{-1}$, for $E_C/D = 0.01$
	 (open squares), $0.25$ (open triangles), and $1$ (dots).
	 Here, $T_K^{\ast}$ is the Kondo temperature for $t\rho = 0.1$.
	 Explicitly, $k_B T_K^{\ast}/D$ equals $3.2 \times 10^{-6}$,
	 $3.6 \times 10^{-5}$, and $6.5 \times 10^{-5}$,
	 for $E_C/D = 0.01$, $0.25$, and $1$, respectively.}
\label{fig:fig4}
\end{figure}

Figure~\ref{fig:fig4} shows the capacitance at the degeneracy points
as a function of temperature. At low temperatures, the capacitance
diverges logarithmically with decreasing $T$, confirming the onset
of the two-channel Kondo effect. Indeed, the two-channel Kondo effect
persists for all values of $E_C/D$ and all values of $t\rho$ up to
perfect transmission, where it breaks down.~\cite{Matveev95} However,
the logarithmic behavior is regained upon further increasing $t\rho$,
marking the reentrance of the two-channel Kondo effect. 

The crossover from the high-temperature to the low-temperature
universal regime is generally marked by the Kondo temperature $T_K$,
which we extract from the Bethe ansatz expression for the slope of the
logarithmically diverging term in the capacitance:~\cite{SS89}
\begin{equation}
C(T) \sim \frac{e^2}{20 k_B T_K} \ln \! \left( T_K/T \right) .
\label{Tk_def}
\end{equation}
Within the two-charge-configuration model of Matveev, $T_K$ is
given by~\cite{Matveev91}
\begin{equation}
T_K = D_{\rm eff}\, t\rho\, \exp
                  \left [
                          -\frac{\pi}{4 t\rho}
                  \right ] ,
\label{Tk_Matveev}
\end{equation}
where $D_{\rm eff}$ is an effective high-energy cutoff, assumed
to be of the order of the charging energy.

In the inset of Fig.~\ref{fig:fig4}, we plotted $T_K/T^*_K$ versus
$(t\rho)^{-1}$, for a wide range of charging energies $E_C/D$.
Here, $T^*_K$ is the Kondo temperature for $t\rho = 0.1$, extracted
separately for each value of $E_C/D$. For weak tunneling, $T_K$
decays exponentially with $(t\rho)^{-1}$, in accordance with
Eq.~(\ref{Tk_Matveev}). In particular, all curves fall on top
of one another in this regime, confirming that $E_C$ enters
the expression for $T_K$ only through the preexponential
factor $D_{\rm eff}$. However, a detailed comparison of the NRG
data and Eq.~(\ref{Tk_Matveev}) reveals an additional dependence
of $D_{\rm eff}$ on $t\rho$. For $E_C/D = 0.01$, for example,
$D_{\rm eff}$ is enhanced by a factor of $4$ upon going from
$t\rho = 0.05$ to $t\rho = 0.2$. At perfect transmission, the
two-channel Kondo effect breaks down, which is signaled
by a cusp in $T_K$ as a function of $t\rho$ [the slope
of the $\ln(T)$ component of $C(T)$ vanishes]. In contrast
to the universality of the weak-tunneling regime, the
position of the cusp in $T_K$ shifts to higher values
of $t\rho$ with increasing $E_C$, showing that one cannot
exclusively parametrize the system by the noninteracting
transmission coefficient of Eq.~(\ref{transmission_coefficient}).

In summary, we devised an NRG approach for solving the
charging of a quantum box connected to a lead by a single-mode
point contact, which uniformly treats all regimes of the
Coulomb blockade. Using this approach we are able to
(i) reveal the rapid breakdown of the perturbation theory,
(ii) obtain a surprising reentrance of the Coulomb-blockade
staircase for large tunneling amplitudes, (iii) confirm
Matveev's scenario for the shape of the charge curve near
perfect transmission, and (iv) obtain two-channel Kondo
behavior at the degeneracy points directly from the
Hamiltonian of Eq.~(\ref{initial_Hamiltonian}).

E.L. and A.S. were supported in part by the Centers of
Excellence Program of the Israel science foundation,
founded by The Israel Academy of Science and Humanities.
F.B.A. was supported in part by DFG grant No. AN 275/2-1.


\begin{thebibliography}{99}
\bibitem{DOS}
        Actually, $\rho$ is the geometric average of the density of
        states in the lead and in the box.
\end{thebibliography}
\end{document}